\documentclass{desyproc}

\begin{document}
\title{Probing QCD Parameters with Top-Quark Data}

\author{{\slshape Sebastian Naumann-Emme}  for the ATLAS and CMS Collaborations\\[1ex]
DESY, Notketra{\ss}e 85, 22607 Hamburg, Germany }

\contribID{xy}  
\confID{7095}
\desyproc{DESY-PROC-2013-XY}
\acronym{TOP2013}
\doi            

\maketitle

\begin{abstract}
  Results from inclusive and differential measurements of the production cross sections for top quarks
  in proton-proton collisions at center-of-mass energies of 7 and 8~TeV are compared
  to predictions at next-to-leading and next-to-next-to-leading order in perturbative Quantum Chromodynamics.
  From these studies,
  constraints on the top-quark mass, the strong coupling constant, and on parton distributions functions are determined.
\end{abstract}

\section{Introduction}

Quantum Chromodynamics (QCD) is the theory of the strong interaction between quarks and gluons.
The only free parameters of the QCD Lagrangian are the quark masses and the strong coupling constant, $\alpha_S$.
The factorization theorem of QCD allows the calculation of cross sections, $\sigma$, to be split 
into hard-scattering matrix elements, $\hat \sigma$, on the one hand and parton distribution functions (PDFs) on the other.
While $\hat \sigma$, describing the short-distance structure of a reaction, is process-dependent but
perturbatively calculable, the PDFs, which account for the non-perturbative long-distance structure, are universal
but have to be determined from experimental data.

The top quark is by far the heaviest of all quarks.
Measurements of the top-quark mass, $m_t$, have been brought to an impressive precision
at the Tevatron, $m_t = 173.20 \pm 0.87$~GeV~\cite{CDF:2013jga}, and at the Large Hadron Collider (LHC),
$m_t = 173.20 \pm 0.95$~GeV~\cite{CMS:2013sfa}.
However, exact relations between these results and theoretically well defined mass schemes have not yet been
established.

The strong coupling constant has been measured in numerous processes and at different energies.
The latest world average, which takes the mass of the Z boson as reference scale,
is $\alpha_{S}(m_{Z}) = 0.1184 \pm 0.0007$~\cite{Beringer:1900zz}.
This average and its remarkable precision are driven by results obtained at relatively low
energies, namely from hadronic decays of $\tau$ leptons and from lattice QCD.
Cross sections for jet production at the LHC allow $\alpha_{S}$ to be probed even up to the TeV scale.
However, the corresponding jet cross sections have typically been calculated only to next-to-leading order (NLO) QCD so far
and they suffer from sizable uncertainties related to choice and variation of the
renormalization and factorization scales, $\mu_R$ and $\mu_F$, as well as from non-perturbative corrections.

PDF groups have released a large number of different PDF sets.
For a given order in perturbation theory, the main differences between these PDF sets arise from the choice of the
included data, the treatment of systematic uncertainties in the data and of correlations, the parametrization at the
starting scale, the chosen heavy-quark scheme, and the values of the quark masses and of $\alpha_{S}(m_{Z})$.
At present, all PDF sets exhibit a significant uncertainty on the gluon density at medium--high parton momentum
fractions, $x$.
This uncertainty affects predictions for Higgs-boson, top-quark, and jet production as well as many scenarios for
new physics beyond the standard model.

In this article, constraints on PDFs, $\alpha_{S}(m_{Z})$, and $m_t$ from LHC top-quark data as well as their interplay
are discussed.
In general, the evolution of such QCD analyses is as follows:
First, identify and potentially maximize the sensitivity of experimental data to the parameters of interest.
Then, understand correlations, both between theory parameters and within the data.
And, eventually, improve PDFs or determine other parameters by including the new data in QCD fits.

\section{Top-Quark Pair Production}

\subsection{The Total Cross Section}

At the LHC, top quarks are produced at relatively high rate,
predominantly in pairs of quarks and anti-quarks  ($t \bar{t}$) from gluon-gluon fusion.
The calculation of the total $t \bar{t}$ cross section, $\sigma_{t \bar{t}}$, to next-to-next-to-leading order (NNLO)
plus next-to-next-to-leading-log (NNLL) resummation has recently been completed~\cite{Czakon:2013goa}.
The uncertainties related to higher orders, estimated via the variation of $\mu_R$ and $\mu_F$,
to the PDFs, to $\alpha_{S}(m_{Z})$, and to $m_t$ now amount to roughly 3\% each.
From the experimental point of view, $\sigma_{t \bar{t}}$ has been measured by the ATLAS and CMS Collaborations
at proton-proton center-of-mass energies, $\sqrt{s}$, of 7 and 8~TeV, using the various $t \bar{t}$ decay channels.
The most precise results have been obtained in the dilepton channel~\cite{Chatrchyan:2012bra,TheATLAScollaboration:2013dja},
both of them yielding a total uncertainty on $\sigma_{t \bar{t}}$ below 5\%.

\begin{figure}[hb]
  \centerline{\includegraphics[width=0.5\textwidth]{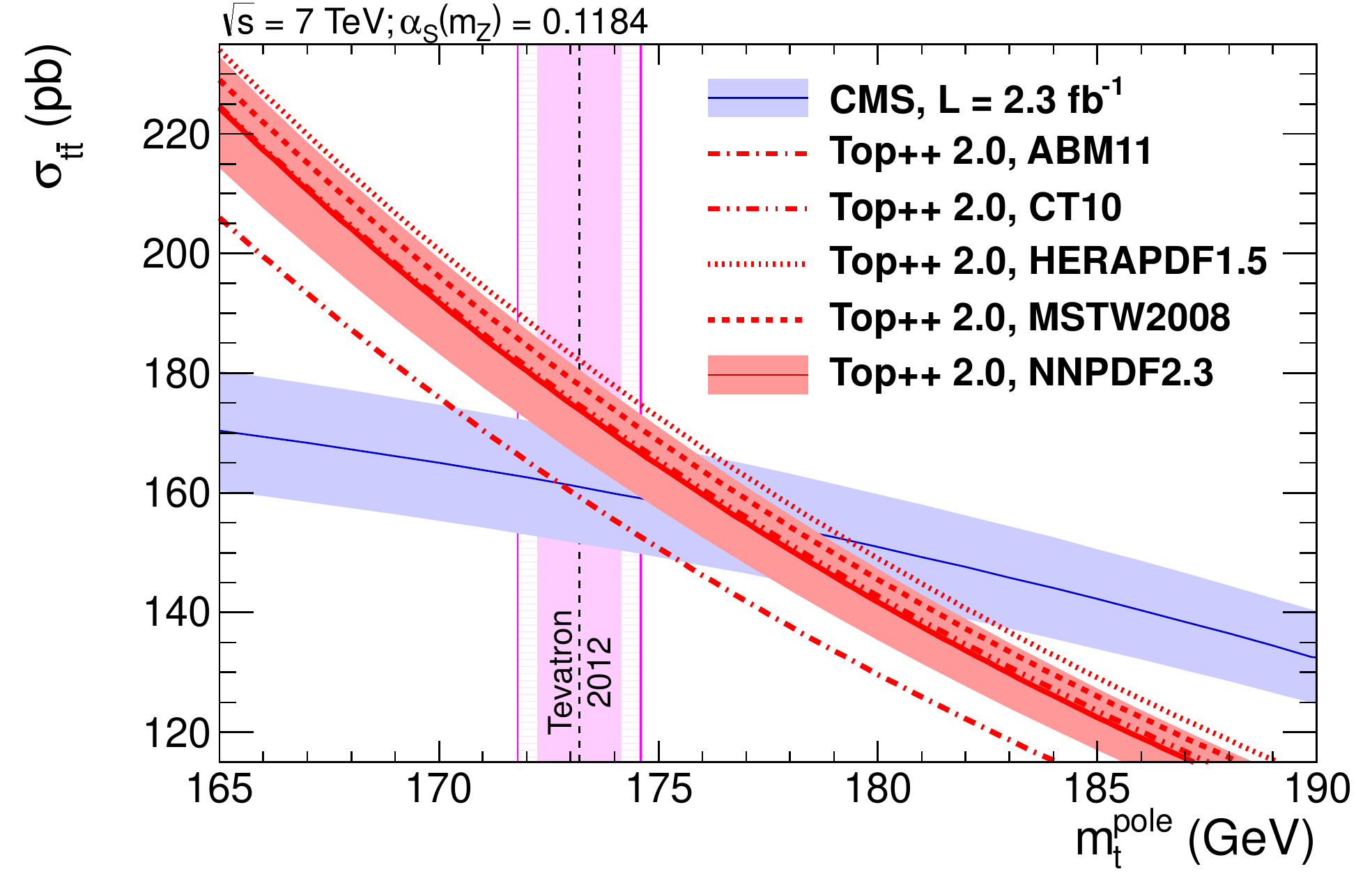}
    \includegraphics[width=0.5\textwidth]{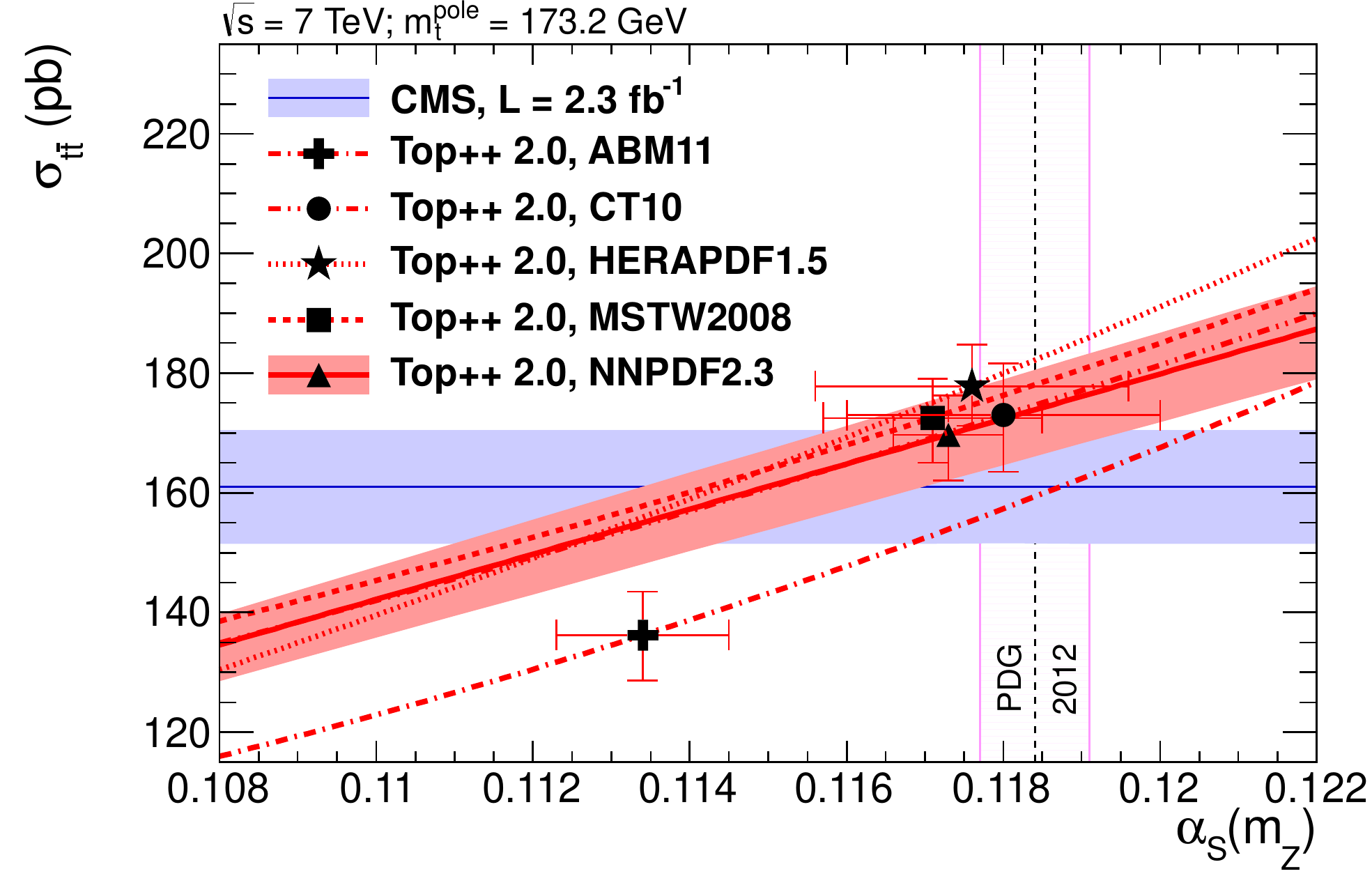}}
  \caption{Predicted $t \bar{t}$ cross section at NNLO+NNLL, as a function of the top-quark mass (left)
    and of the strong coupling constant (right),
    using five different NNLO PDF sets, compared to the cross section measured by CMS~\cite{Chatrchyan:2013haa}.
    \label{Fig1}}
\end{figure}

The predicted $\sigma_{t \bar{t}}$ strongly depends on the assumed values of $m_t$ and $\alpha_{S}(m_{Z})$,
but also the measured cross section can depend on them.
Dependencies of the measured cross section arise from the acceptance corrections,
which are derived using simulated $t \bar t$ events.
Figure~\ref{Fig1} compares CMS' most precise single measurement of $\sigma_{t \bar{t}}$ \cite{Chatrchyan:2012bra},
which was obtained at $\sqrt{s} = 7$~TeV, to the NNLO+NNLL prediction with five different NNLO PDF sets.
These PDF sets are provided for a series of $\alpha_{S}(m_{Z})$ values,
which allows the full correlation between the choice of $\alpha_{S}(m_{Z})$ and the parton densities to be preserved.
Relatively small differences are found between four of the five PDF sets,
namely between CT10, HERAPDF1.5, MSTW2008, and NNPDF2.3.
ABM11, by contrast, does not only have a smaller default value of $\alpha_{S}(m_{Z})$
but also a smaller gluon density, which results in a lower $\sigma_{t \bar{t}}$ prediction compared to the other PDF sets
at any given $\alpha_{S}(m_{Z})$ value.
While the measured $\sigma_{t \bar{t}}$ has a sizable $m_t$ dependence,
only a minor dependence on $\alpha_{S}(m_{Z})$ was found.

\begin{figure}[hb]
  \centerline{\includegraphics[width=0.5\textwidth]{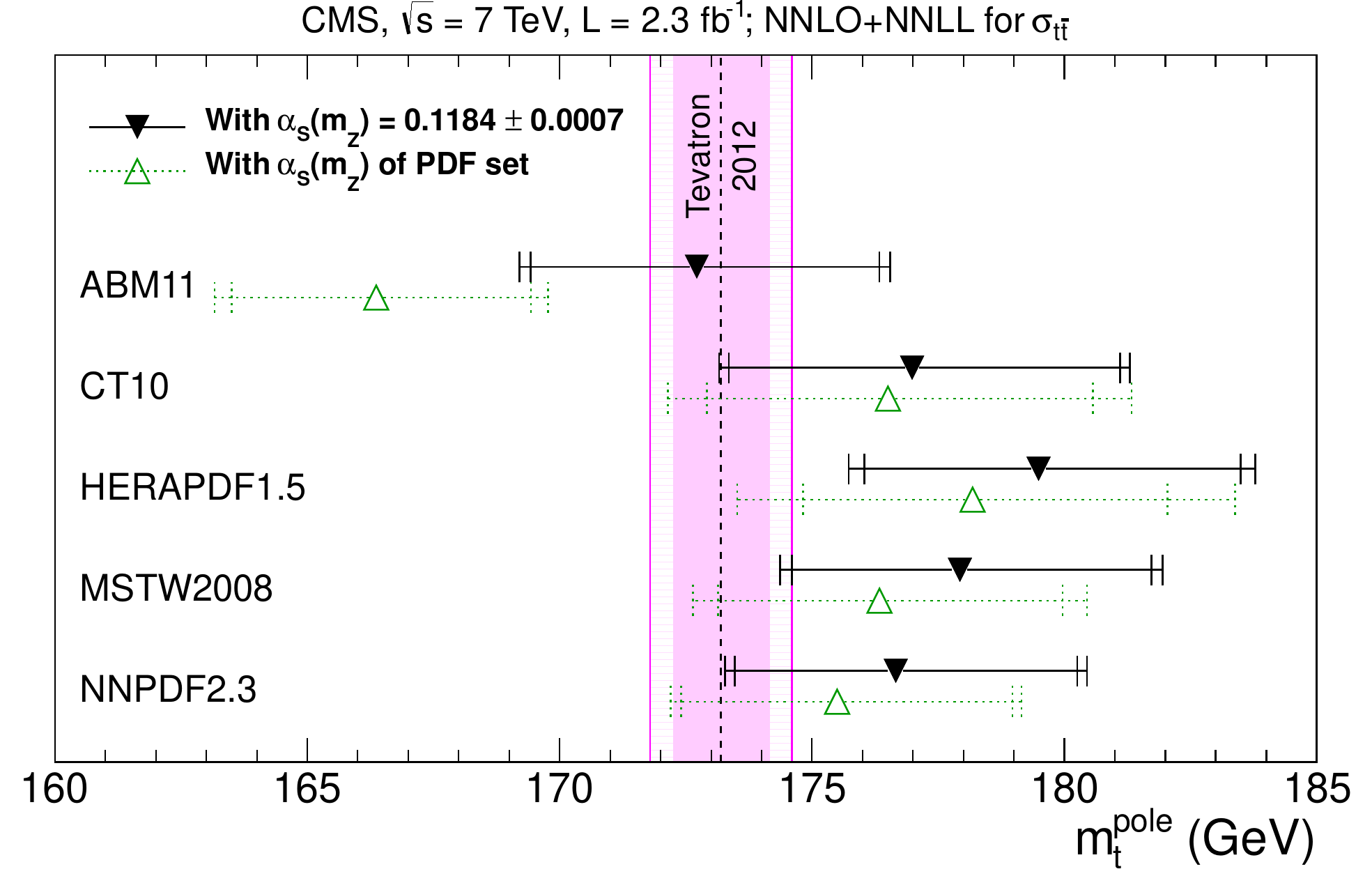}
    \includegraphics[width=0.5\textwidth]{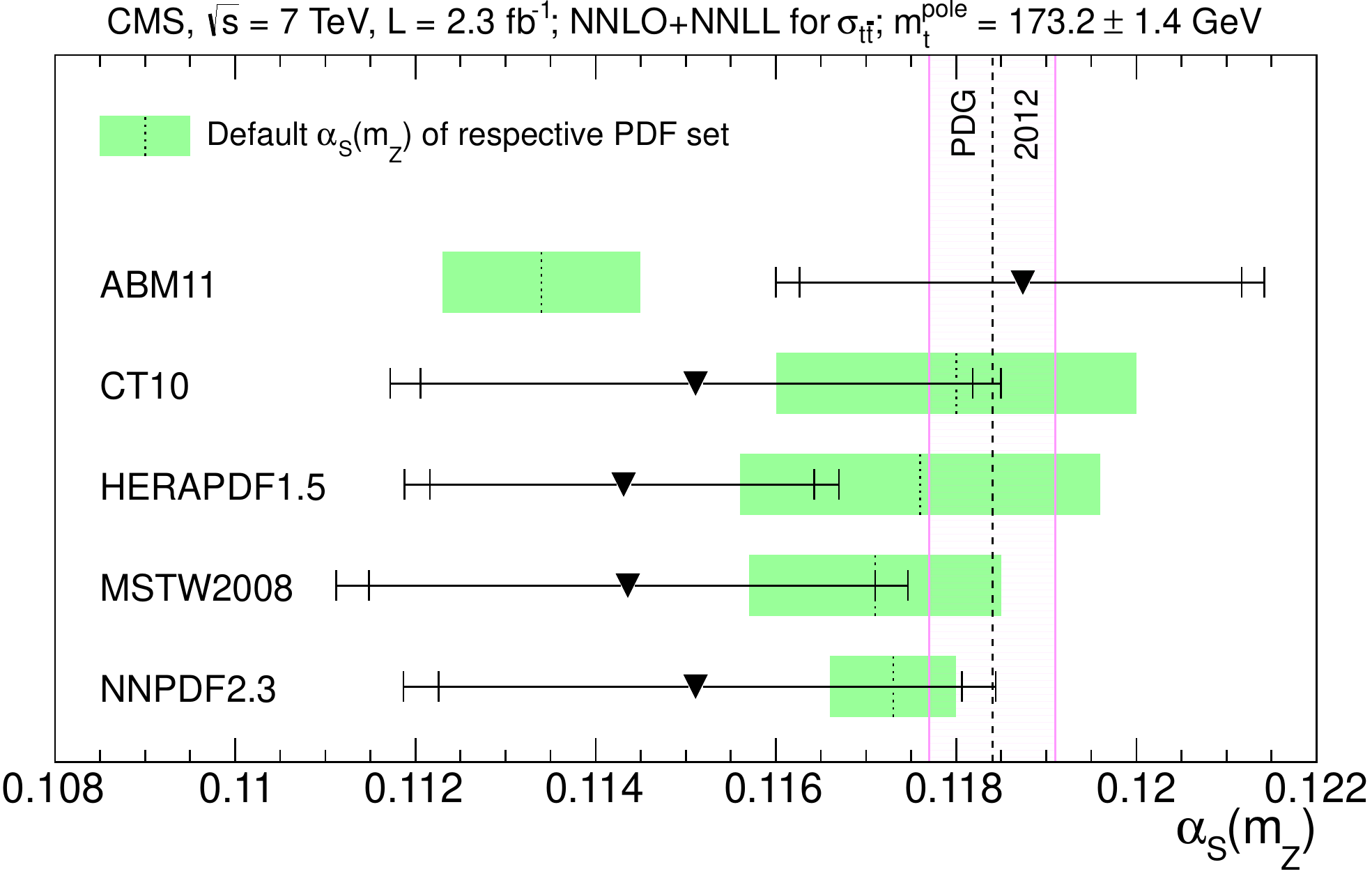}}
  \caption{Results obtained for the top-quark mass (left) and for the strong coupling constant (right)
    by comparing the measured $t \bar{t}$ cross section from CMS
    to the prediction at NNLO+NNLL using five different NNLO PDF sets~\cite{Chatrchyan:2013haa}.
    \label{Fig2}}
\end{figure}

CMS used this comparison between measured and predicted $\sigma_{t \bar{t}}$ for extractions of $m_t$ and
$\alpha_{S}(m_{Z})$~\cite{Chatrchyan:2013haa}.
The NNLO+NNLL prediction was taken as a Bayesian prior to the cross-section measurement,
which enabled the construction of marginalized posteriors in $m_t$ and $\alpha_{S}(m_{Z})$.
The measured cross section was parametrized using a Gaussian probability function along $\sigma_{t \bar{t}}$.
The PDF uncertainty on the predicted $\sigma_{t \bar{t}}$ was also assumed to be Gaussian and convoluted
with a step function that yields equal probabilities for all $\sigma_{t \bar{t}}$ values covered by the
$\mu_R$ and $\mu_F$ variations and vanishes elsewhere.
No big changes were found when trying different parametrizations for the scale uncertainty.
The marginalized posteriors were then obtained by integrating over $\sigma_{t \bar{t}}$.
For given values of $\alpha_{S}(m_{Z})$ or $m_t$, these posteriors yield the most probable
$m_t$ and $\alpha_{S}(m_{Z})$ values, respectively, together with Bayesian confidence intervals
that account for the uncertainty on the measured cross section and the PDF and scale uncertainties on the
predicted cross section. Additionally, the following uncertainties were taken into account:
\begin{itemize}
\item An uncertainty of 0.65\% on the LHC beam energy ($E_{\text{LHC}}$)~\cite{Wenninger:2013},
  translating into $46$~GeV on the nominal $\sqrt{s}$ value of 7~TeV.
\item For the $m_t$ determination, the uncertainty of $0.0007$ on the $\alpha_S (m_{Z})$ world average, which was used
  as constraint.
\item For the $m_t$ determination, an uncertainty of 1~GeV on the equality of top-quark pole mass and the top-quark mass
  in the Monte Carlo simulation ($m_t^{\text{MC}}$)~\cite{Buckley:2011ms},
  since the simulation was employed for the acceptance corrections in the $\sigma_{t \bar{t}}$ measurement.
\item For the $\alpha_S (m_{Z})$ determination, an uncertainty of 1.4~GeV on the Tevatron average for $m_t$, which was used
  as constraint. This variation accounts for both the 0.9~GeV uncertainty of the Tevatron average itself and the 1~GeV
  uncertainty in relating $m_t^{\text{MC}}$, employed also to calibrate these direct mass measurements, to the top-quark
  pole mass.
\end{itemize}
Using NNPDF2.3, the results are
\begin{eqnarray*}
  m_t &=& 176.7^{+3.1}_{-2.8}(\text{exp.}){}^{+1.5}_{-1.3}(\text{PDF}){}^{+0.9}_{-0.9}(\text{scale}){}^{+0.7}_{-0.7}(\alpha_{S}){}^{+0.9}_{-0.9}(E_{\text{LHC}}){}^{+0.5}_{-0.4}(m_t^{\text{MC}}) \ \text{GeV} \\
  &=& 176.7^{+3.8}_{-3.4} \ \text{GeV}
\end{eqnarray*}
and, alternatively,
\begin{eqnarray*}
  \alpha_{S}(m_{Z}) &=& 0.1151^{+0.0025}_{-0.0025}(\text{exp.}){}^{+0.0013}_{-0.0011}(\text{PDF}){}^{+0.0009}_{-0.0008}(\text{scale}){}^{+0.0013}_{-0.0013}(m_t){}^{+0.0008}_{-0.0008}(E_{\text{LHC}}) \\
  &=& 0.1151^{+0.0033}_{-0.0032}.
\end{eqnarray*}
The results with all five PDF sets are shown in Figure~\ref{Fig2}.
These are the first extractions of the top-quark pole mass at full NNLO QCD, of $\alpha_{S}(m_{Z})$ from top-quark data,
and of $\alpha_{S}(m_{Z})$ at full NNLO QCD from a hadron collider.
There are only small differences between the result obtained with CT10, HERAPDF1.5, MSTW2008, and NNPDF2.3,
while the smaller gluon density of ABM11 requires either a lower $m_t$ or a higher $\alpha_{S}(m_{Z})$ value
to reproduce the $\sigma_{t \bar{t}}$ measured by CMS.
Using ABM11 with its default $\alpha_S (m_Z)$ of $0.1134 \pm 0.0011$ would yield $m_t = 166.3^{+3.3}_{-3.1}$~GeV,
which is significantly lower than the results from direct $m_t$ measurements
and than the results obtained via $\sigma_{t \bar{t}}$ when using the other PDF sets.

First studies illustrating the impact of the total $\sigma_{t \bar{t}}$ as measured at LHC and Tevatron
in particular on the gluon PDF have been released by different authors~\cite{Beneke:2012wb,Czakon:2013tha,Alekhin:2013nda}.
However, more work is needed to accurately incorporate all systematic uncertainties and correlations, both between
the PDFs, $\alpha_S$, and $m_t$ and among the experimental data, as well as the experimental $m_t$ dependencies.
Ratios of the $t \bar t$ cross section measured at different center-of-mass energies (8 to 7~TeV; later: 14 to 8~TeV) also
have promising prospects for PDF fits, since the PDF uncertainties on the predicted ratios are significantly
larger than the combined $\mu_R$, $\mu_F$, $\alpha_S$, and $m_t$ uncertainties~\cite{Czakon:2013tha},
but such cross-section ratios require a thorough understanding of the correlations between the systematic uncertainties
on the measured $\sigma_{t \bar{t}}$ at the different center-of-mass energies.

\clearpage

\subsection{Differential Cross Sections}

ATLAS and CMS have measured a variety of (normalized) differential cross sections for $t \bar t$
production \cite{Aad:2012hg,Chatrchyan:2012saa,CMS:fxa,CMS:cxa,TheATLAScollaboration:2013eja}.
These results, discussed in more detail in~\cite{Aldaya:TOP2013},
can be compared to predictions at NLO or, in some cases (namely the distributions as a function of the
transverse momentum and the rapidity of the top-quarks as well as the invariant mass of the $t \bar t$ system),
to calculations at approximate NNLO.

\begin{figure}[hb]
  \centerline{\includegraphics[width=0.5\textwidth]{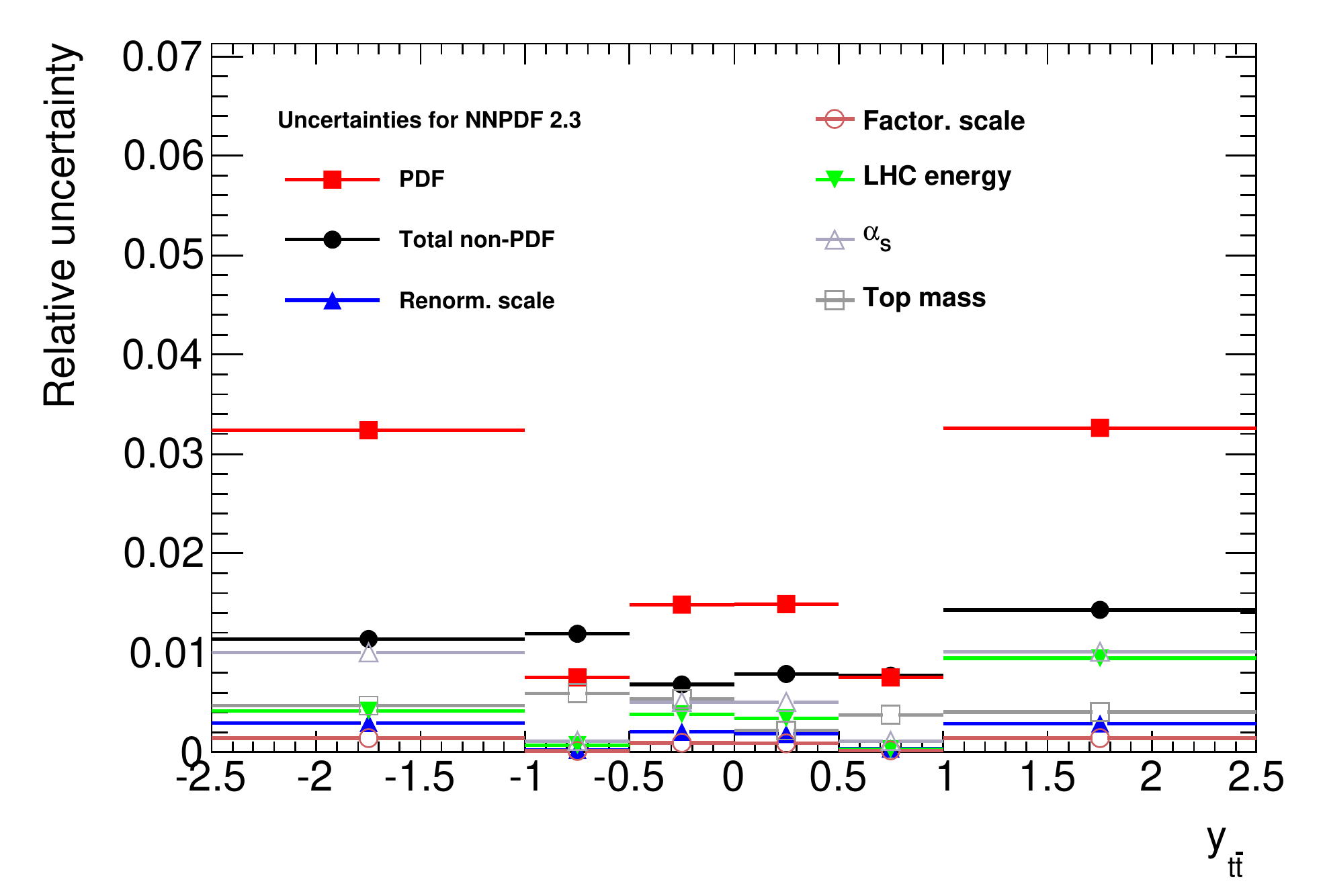}
    \includegraphics[width=0.5\textwidth]{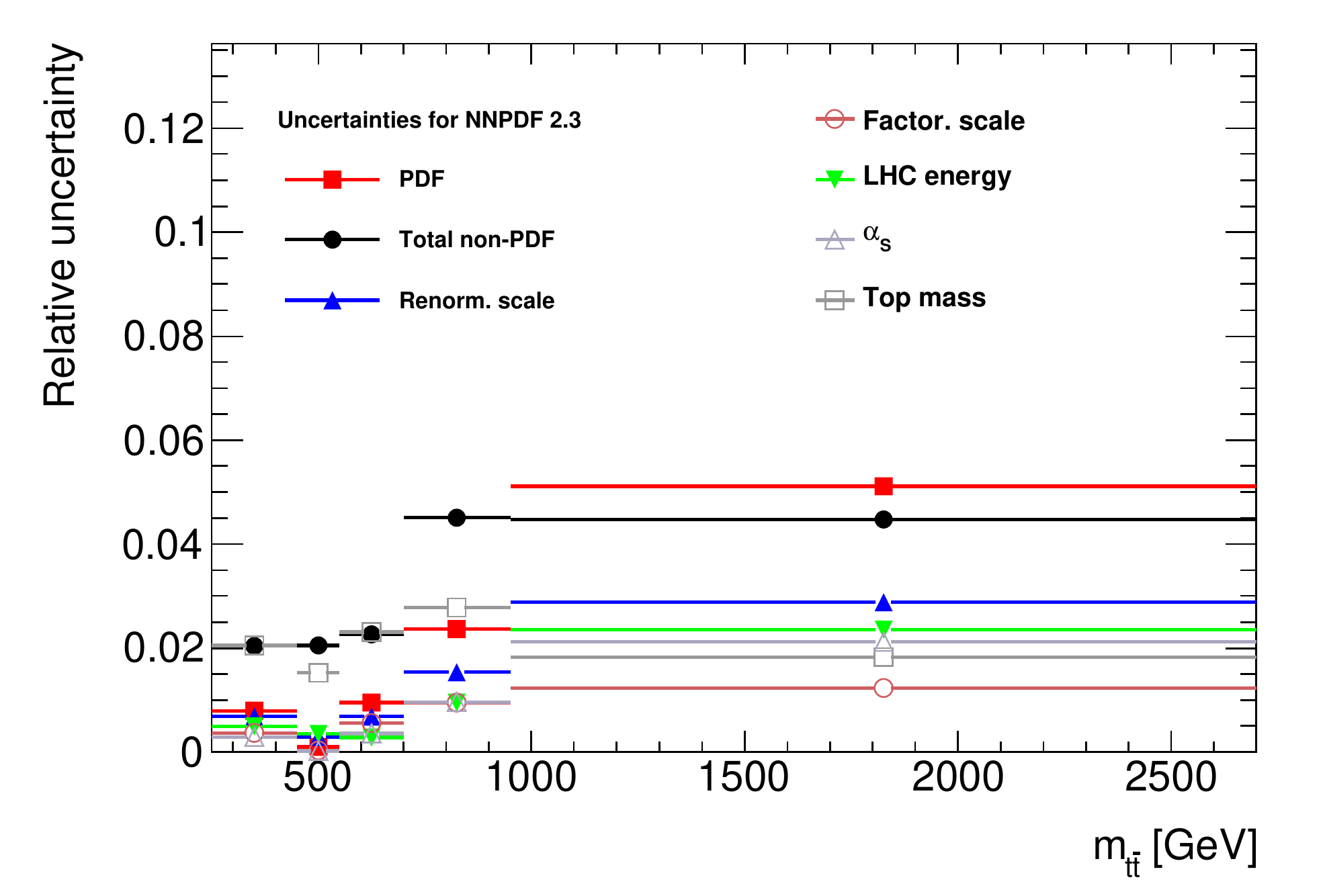}}
  \caption{Relative uncertainties on the NLO prediction 
    with NNPDF2.3 for the normalized $t \bar t$ cross section
    as a function of the rapidity (left) and the invariant mass (right) of the $t \bar t$
    system~\cite{TheATLAScollaboration:2013nn}.
    \label{Fig3}}
\end{figure}

In general, kinematic regions in which the PDF uncertainty on the predicted cross section
is larger than other modeling uncertainties are considered to have
the largest potential to improve the accuracy of future PDF fits.
The ATLAS Collaboration compared differential $t \bar t$ cross sections at $\sqrt{s} = 7$~TeV
to predictions at NLO QCD 
with different NLO PDF sets~\cite{TheATLAScollaboration:2013nn,TheATLAScollaboration:2013eja}.
The best PDF sensitivity was found in the rapidity and the invariant mass of $t \bar t$ system, $y_{t \bar t}$ and $m_{t \bar t}$.
The size of the corresponding theory uncertainties are illustrated in Figure~\ref{Fig3}.
Both $y_{t \bar t}$ and $m_{t \bar t}$ are directly correlated with the momenta of the incoming partons.
Large rapidities require one incoming parton with high $x$, the other one with small $x$.
Large $m_{t \bar t}$ values also probe the high-$x$ regime.
However, it has to be kept in mind that electroweak corrections to differential $t \bar t$ cross section
are known to be non-negligible, in particular for high transverse momenta and invariant masses but also for the
shape of the $y_{t \bar t}$ distribution (as discussed, for example, in~\cite{Kuhn:2013zoa}),
and that these corrections are typically not yet included in these comparisons.

Using the first 2.1~fb${}^{-1}$ of data at $\sqrt{s} = 7$~TeV, ATLAS quantified the compatibility between measured
differential cross sections in the lepton+jets channel
and the corresponding NLO predictions 
with five different NLO PDF sets
(ABM11, CT10, HERAPDF1.5, MSTW2008, NNPDF2.3), taking into account all experimental and theoretical uncertainties
as well as bin-to-bin correlation for the data~\cite{TheATLAScollaboration:2013nn}.
The best separation strength was found for $y_{t \bar t}$,
where the $\chi^2$ probabilities range from 21\% for CT10 to 83\% for NNPDF2.3.

\begin{figure}[hb]
  \centerline{\includegraphics[width=0.33\textwidth]{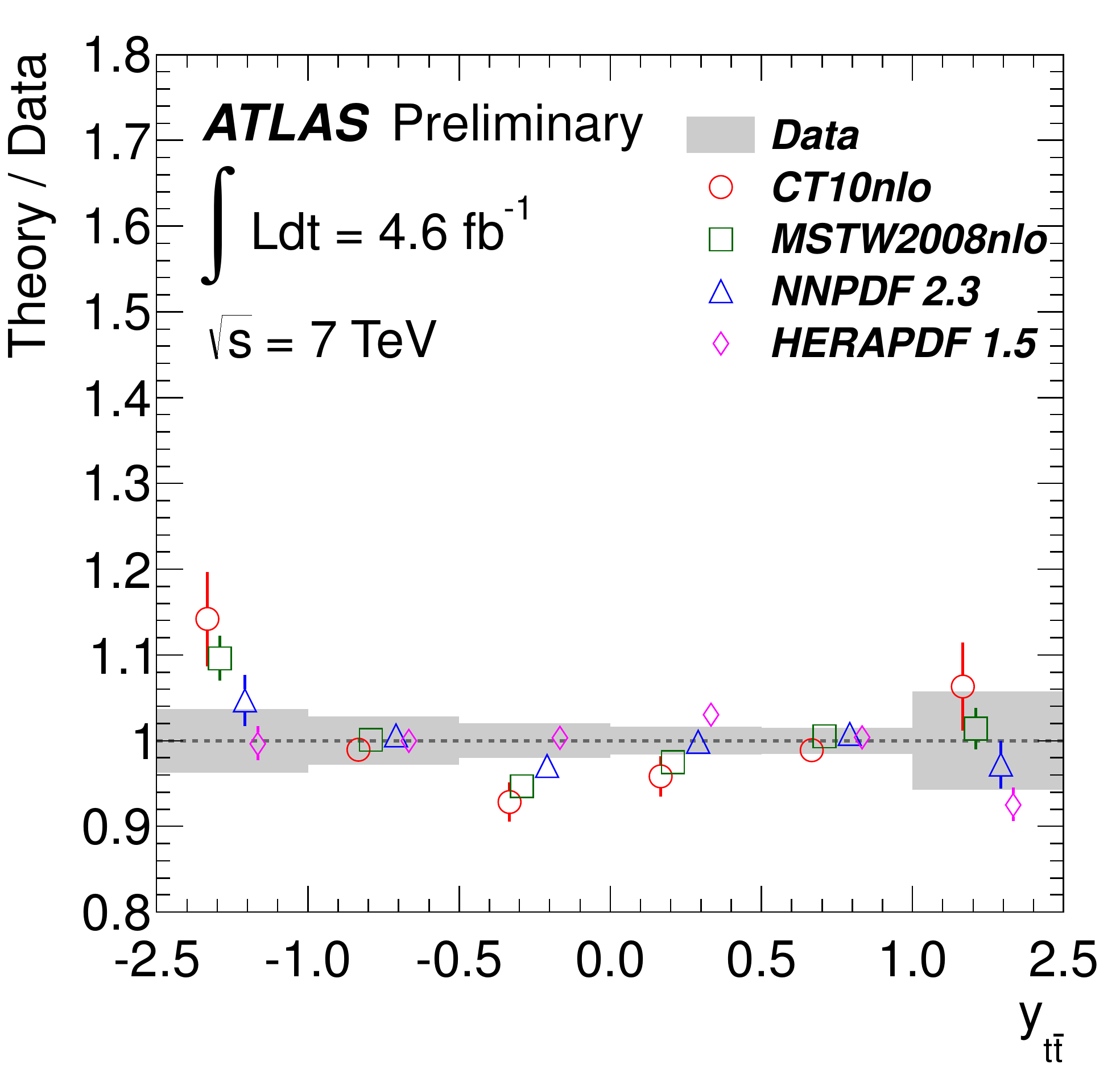}
    \includegraphics[width=0.33\textwidth]{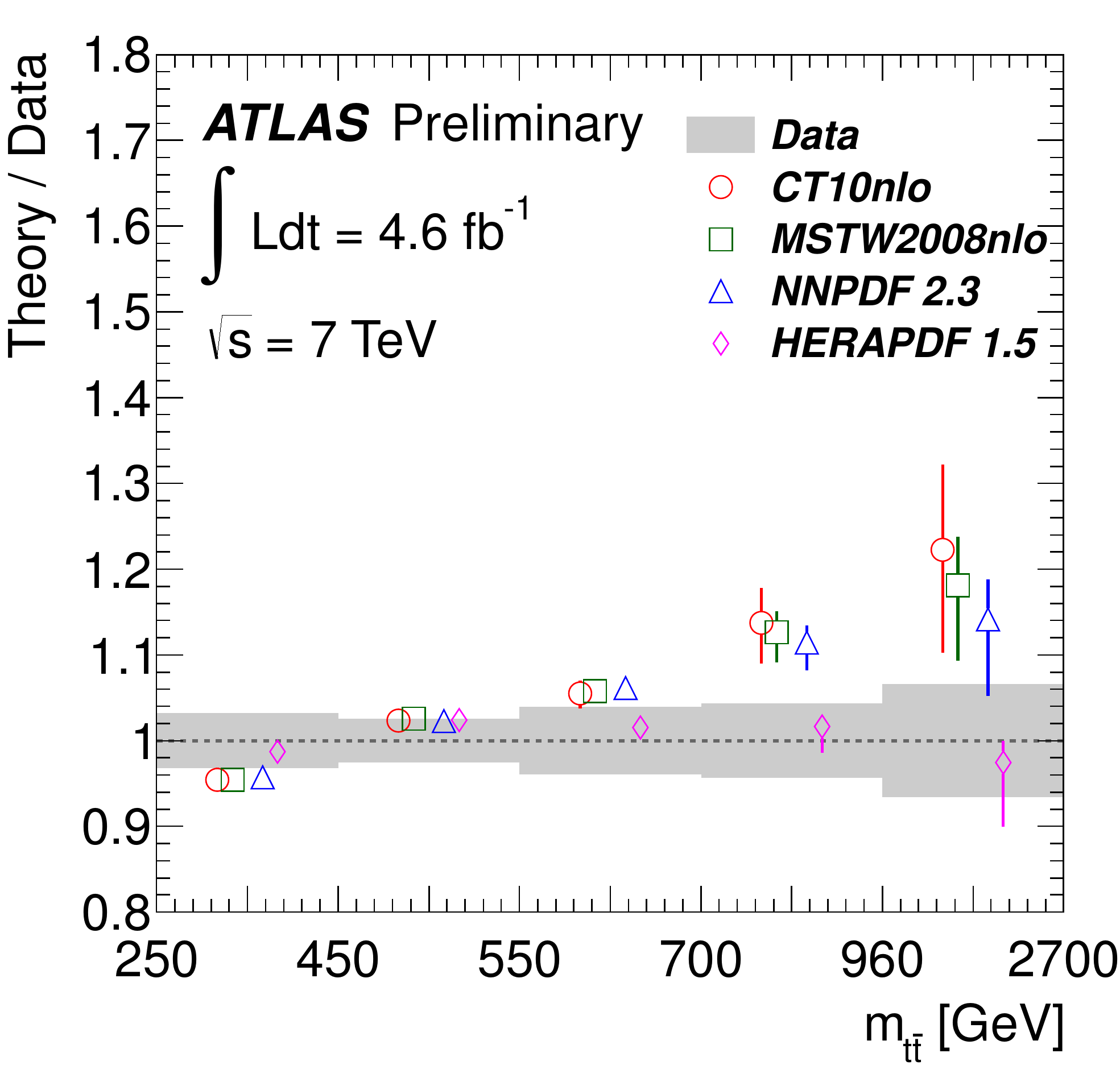}
    \includegraphics[width=0.33\textwidth]{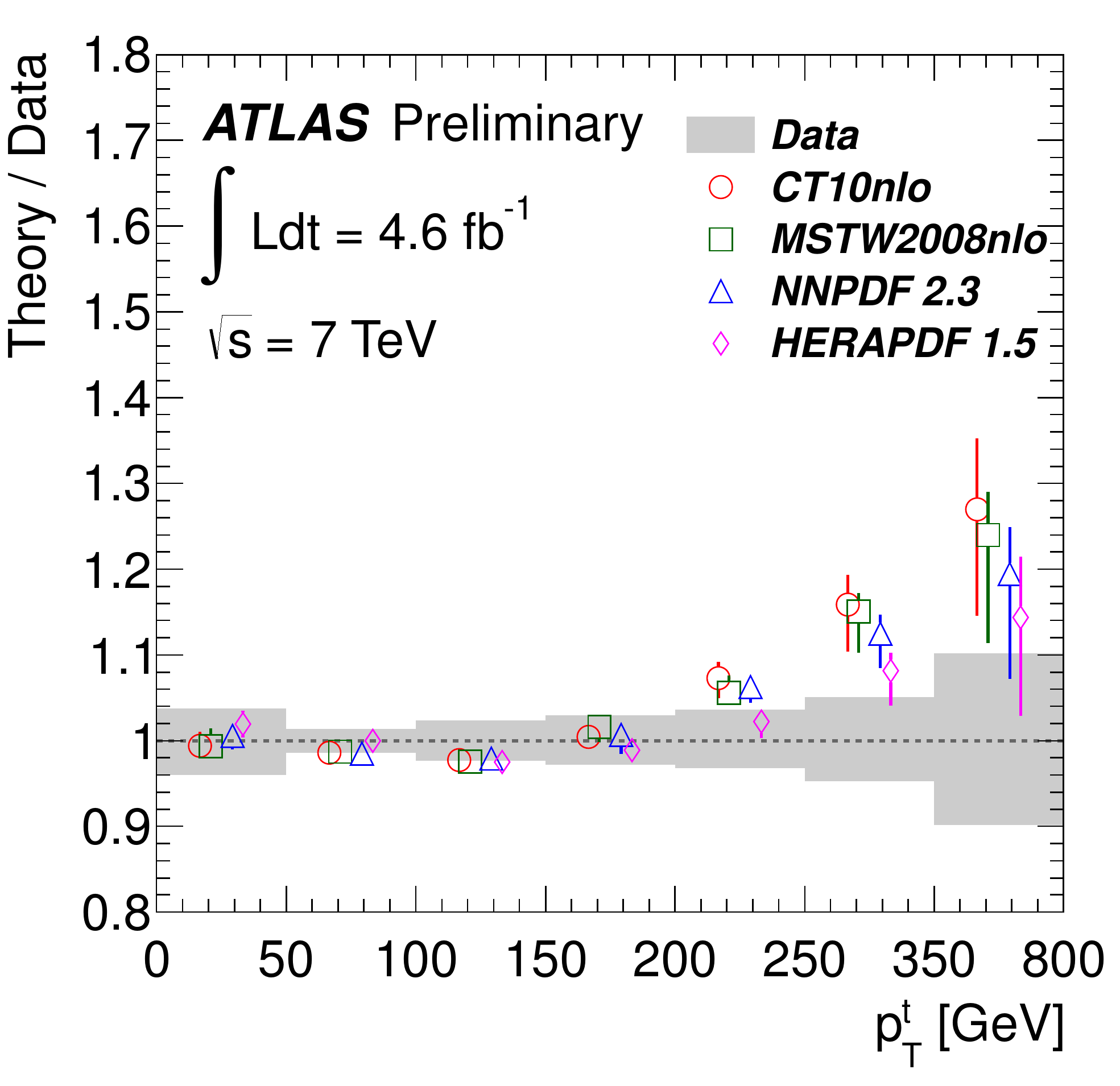}}
  \caption{Ratios of the NLO predictions 
    with four different NLO PDF sets to the measured normalized $t \bar t$ cross section
    as a function of rapidity (left) and invariant mass (center) of the $t \bar t$ system as well
    as the transverse momentum of the top quarks (right) \cite{TheATLAScollaboration:2013eja}.
    \label{Fig4}}
\end{figure}

Recently, the ATLAS Collaboration released updated results for differential $t \bar t$ cross
sections~\cite{TheATLAScollaboration:2013eja}, now using the full dataset at 7~TeV, corresponding to 4.6~fb${}^{-1}$.
The compatibility between data and NLO prediction is shown in Figure~\ref{Fig4} for
$y_{t \bar t}$, $m_{t \bar t}$, and the transverse momentum of the top quarks, $p_{T,t}$.
In all three cases, a significant tension in shape between the predictions with the various PDF sets can be seen.
As before, the level of agreement between data and prediction appears to be better with MSTW2008 than with CT10
and better with NNPDF2.3 compared to MSTW2008.
However, the new data seems to prefer the prediction with HERAPDF1.5.
Again, electroweak corrections are not yet included here but could yield a non-negligible contribution.

\section{Production of Single Top Quarks}

The production of single top quarks occurs via weak, charged-current interactions.
At the LHC, the dominant process is the t-channel exchange of a virtual W boson between
a light quark from one of the colliding protons and a bottom quark from the other proton.
Since the up-quark density in protons is about twice as high as the down-quark density,
the cross section for the production of single top quarks is about twice as high as the
cross section for single anti-top quarks.
Precise measurements of the ratio $R_t = \frac{\sigma_t}{\sigma_{\bar t}}$ can provide
a handle on the ratio of the $u$/$d$ densities in the proton.
They probe the kinematic regime $0.02 \lesssim x \lesssim 0.5$ and are thus
complementary to measurements via the charge asymmetry in $W$-boson production,
which probe $0.001 \lesssim x \lesssim 0.1$ at the LHC and $0.005 \lesssim x \lesssim 0.3$ at the Tevatron.

\begin{figure}[htp]
  \centerline{\includegraphics[width=0.5\textwidth]{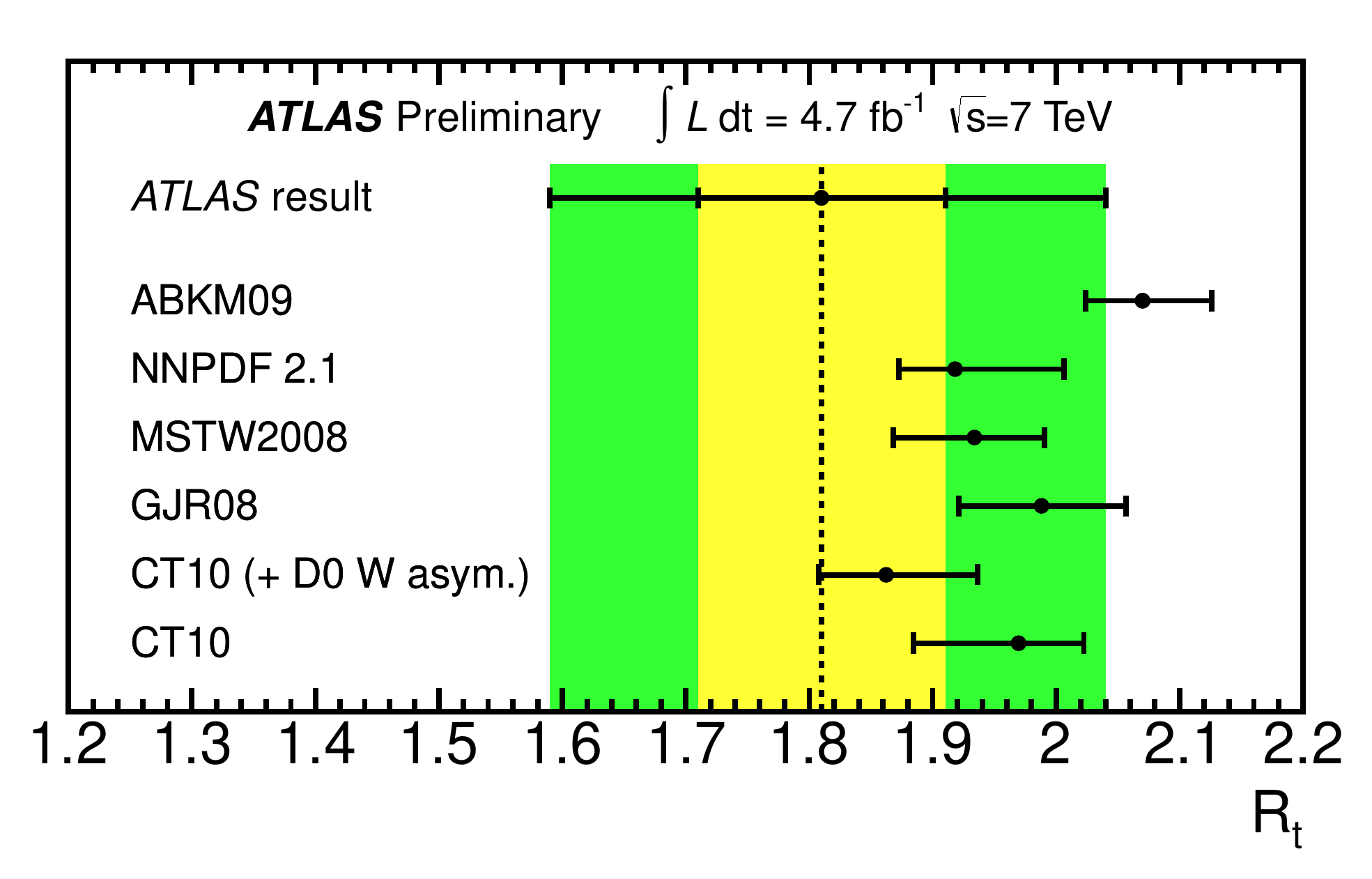}
    \includegraphics[width=0.5\textwidth]{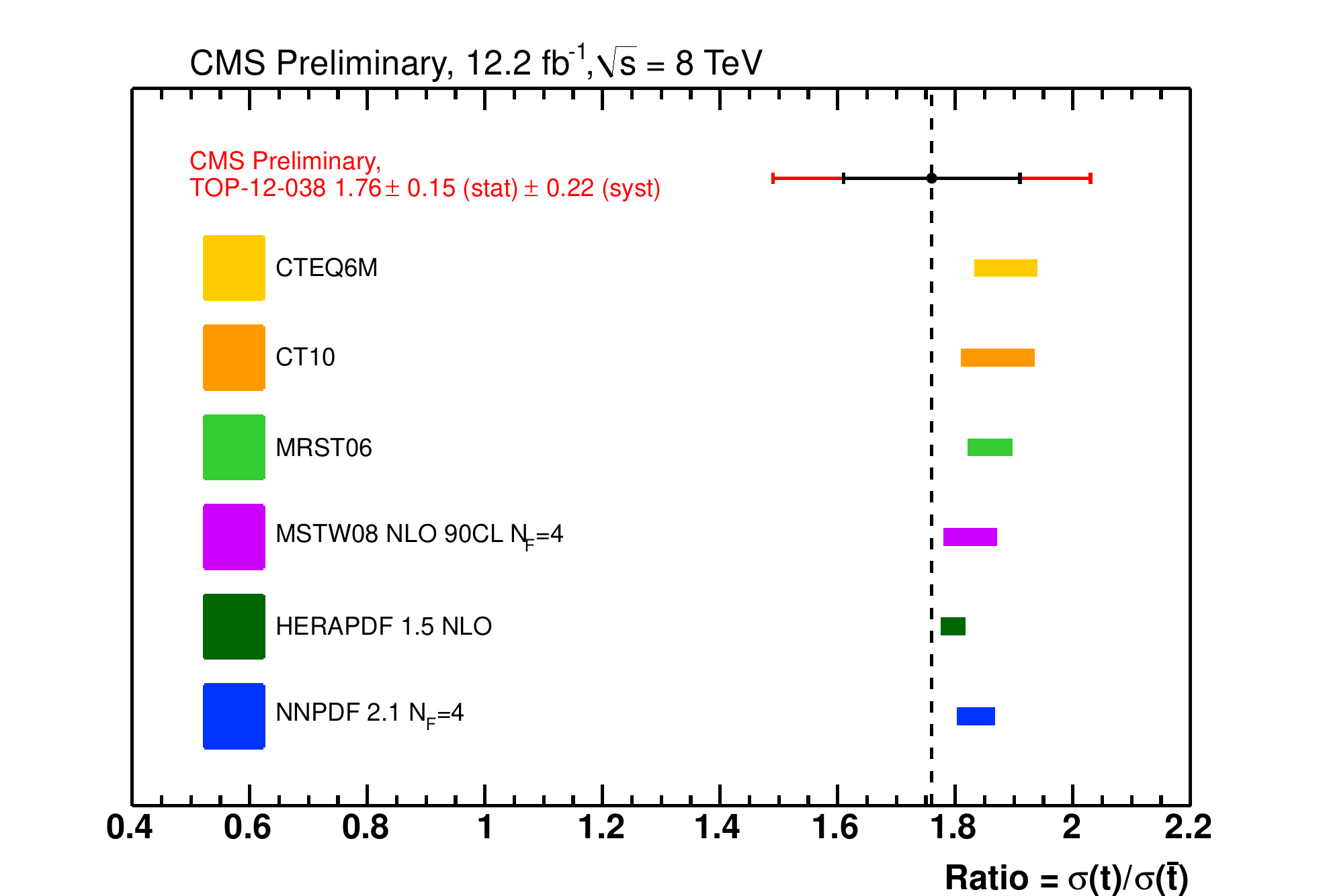}}
  \caption{Comparisons between the ratio of the production cross sections for single top quarks and single anti-top quarks,
    measured by ATLAS at $\sqrt{s}$ = 7~TeV (left)~\cite{ATLAS:2012tla}
    and by CMS at 8~TeV (right)~\cite{CMS:jwa}, and the NLO prediction with various PDF sets.
    \label{Fig5}}
\end{figure}

The ratio $R_t$ has been measured by ATLAS~\cite{ATLAS:2012tla} and CMS~\cite{CMS:jwa}
at $\sqrt{s}$ = 7~TeV and 8~TeV, respectively, to be:
\begin{eqnarray*}
  R_t \ (\text{7 TeV}) &=& 1.81 \pm 0.10 \ (\text{stat.}) \ {}^{+0.21}_{-0.20} \ (\text{syst.}) = 1.81^{+0.23}_{-0.22}, \quad \text{and} \\
  R_t \ (\text{8 TeV}) &=& 1.76 \pm 0.15 \ (\text{stat.}) \ \pm 0.22 \ (\text{syst.}) = 1.76 \pm 0.27.
\end{eqnarray*}
In both cases, the sign of the top-quark charge was inferred from the reconstructed charge of the final-state lepton that
had been associated to the top-quark decay.
The observed $R_t$ are compatible with the predictions at NLO QCD.
This is shown in Figure~\ref{Fig5} using predictions with various PDF sets.
The spread of the predictions with different PDF sets is approximately of the same size as the uncertainty on the
predictions. 
Apart from the light-quark PDFs and the renormalization and factorization scales,
the predicted cross sections for single top-quark production depend also on the choice of the heavy-flavor scheme
(fixed-flavor schemes with four or five active flavors versus variable flavor schemes), the bottom-quark density
in the proton, and the bottom-quark mass.
However, the uncertainty on the measured $R_t$ is currently still roughly more than twice
as large as the total uncertainty on the prediction.

\section{Conclusions}

The large samples of top-quark data that are being collected at the LHC enable new and ever more precise QCD analyses.

The precisely measured total cross section for $t \bar t$ production together with the prediction at NNLO+NNLL QCD
allows for extractions of the top-quark pole mass that are significantly more precise than
previous determinations of the top-quark mass from cross sections.
Alternatively, when constraining $m_t$ to the average of previous measurements,
the $t \bar t$ cross section enables the first $\alpha_S (m_Z)$ determination at NNLO QCD at a hadron collider.
The precision is competitive with other $\alpha_S (m_Z)$ measurements.
Furthermore, the inclusive $t \bar t$ cross section is currently the only process that directly allows the high-$x$
gluon PDF to be probed at full NNLO QCD.
An improved precision on the gluon PDF is crucial not only for future top-quark analyses but also many Higgs-boson analyses
and new-physics searches.
Differential $t \bar t$ cross sections are starting to allow for even more explicit PDF discrimination.
The most sensitive distributions are the differential cross sections as a function of the rapidity and
the invariant mass of the $t \bar t$ system.
In any of these QCD analyses using $t \bar t$ cross sections, it is imperative to consider the full correlations
between $m_t$, $\alpha_S$, and the gluon PDF as well as the correlations within the experimental data.

A handle on the ratio of the $u$-quark to $d$-quark PDFs can eventually be obtained from
more precise measurements of the charge ratio in t-channel production of single top quarks.

\clearpage
 

\begin{footnotesize}

\end{footnotesize}


\end{document}